\documentclass[twocolumn, showpacs, preprintnumbers, amsmath, amssymb, superscriptaddress, prl]{revtex4}

\usepackage{graphicx}
\usepackage{dcolumn}
\usepackage{bm}
\usepackage{amssymb}

\def\ket#1{| {#1} \rangle}

\begin{document}

\title{CNOT and Bell-state analysis in the weak-coupling cavity-QED regime}

\author{Cristian Bonato}
\affiliation{Huygens Laboratory, Leiden University, P.O. Box 9504, 2300 RA Leiden, the Netherlands}

\author{Florian Haupt}
\affiliation{University of California Santa Barbara, Santa Barbara, California 93106, USA}

\author{Sumant S.R. Oemrawsingh}
\affiliation{Huygens Laboratory, Leiden University, P.O. Box 9504, 2300 RA Leiden, the Netherlands}

\author{Jan Gudat}
\affiliation{Huygens Laboratory, Leiden University, P.O. Box 9504, 2300 RA Leiden, the Netherlands}

\author{Dapeng Ding}
\affiliation{Huygens Laboratory, Leiden University, P.O. Box 9504, 2300 RA Leiden, the Netherlands}

\author{Martin P. van Exter}
\affiliation{Huygens Laboratory, Leiden University, P.O. Box 9504, 2300 RA Leiden, the Netherlands}

\author{Dirk Bouwmeester}
\affiliation{Huygens Laboratory, Leiden University, P.O. Box 9504, 2300 RA Leiden, the Netherlands}
\affiliation{University of California Santa Barbara, Santa Barbara, California 93106, USA}

\begin{abstract}
We propose an interface between the spin of a photon and the spin of an electron confined in a quantum dot embedded in a microcavity operating in the weak coupling regime. This interface, based on spin selective photon reflection from the cavity, can be used to construct a CNOT gate, a multi-photon entangler and a photonic Bell-state analyzer. Finally, we analyze experimental feasibility, concluding that the schemes can be implemented with current technology.
\end{abstract}


\maketitle

Hybrid quantum information systems hold great promises for the development
of quantum communication and computing since they allow exploiting
different quantum systems at the best of their potentials. For example, in
order to build a quantum network \cite{ciracPRL97}, photons are excellent candidates for
long-distance transmission while quantum states of matter are preferred
for local storage and processing. Hybrid (photon-matter) systems can also
be used to effectively enable strong nonlinear interactions between single
photons \cite{duanPRL04, schoenPRA07, devittPRA07}. Several
systems have been identified as candidates for local matter
qubits, for example atoms \cite{blinovNature09, weberPRL09}, ions \cite{blattWinelandNature08}, superconducting circuits \cite{walraffNat04, ansmannNature09}, and
semiconductor quantum dots \cite{imamogluPRL99, calarcoPRA03, hansonRMP07}, and their coupling strengths to optical modes
have been investigated.\\
Quantum information protocols based on cavity-QED often require the system to operate in the strong-coupling regime \cite{monroeNature02, duanPRL04, huPRB08_faraday, huPRB08_entangler}, where the vacuum Rabi frequency of the dipole $g$ exceeds both the cavity and dipole decay rates. However, in the bad cavity limit, where $g$ is smaller than the cavity decay rate, the coupling between the radiation and the dipole can drastically change the cavity reflection and transmission properties \cite{fanOL05, waksPRL06, auffevesPRA07}, allowing quantum information schemes to operate in the weak coupling regime. We exploit this regime, using spin selective dipole coupling, for a system consisting of a single electron charged self-assembled GaAs/InAs quantum dot in a micropillar resonator \cite{reitzensteinAPL09, stoltzAPL05}. The potential of this system has also been recognized in \cite{huPRB09}. We first show that this specific system can lead to a quantum CNOT gate with the confined electron spin as control qubit and the incoming photon spin as target qubit. We apply the CNOT gate to generate multi-photon entangled states. We then construct a complete two-photon Bell-state analyzer (BSA). Complete deterministic BSA is an important prerequisite for many quantum information protocols like superdense coding, teleportation or entanglement swapping. It cannot be performed with linear optics only \cite{calsamigliaAPB01}, while it can be done using nonlinear optical processes \cite{kimPRL01} (with low efficiency) or employing measurement-based nonlinearities in nondeterministic schemes \cite{KLMnature00}. Deterministic complete BSA has been shown in a scheme which is conceptually different from the one presented here, exploiting entanglement in two or more degrees of freedom of two photons \cite{kwiatPRA98, walbornPRA03}. We conclude with a discussion on the experimental feasibility of the proposed schemes.

In the limit of a weak incoming field, a cavity with a dipole behaves like a
linear beamsplitter whose reflection ($r$) and transmission ($t$) are given by \cite{auffevesPRA07}:
\begin{equation}
r = \frac{i\xi}{1+i\xi} \qquad t= -\frac{1}{1+i\xi}, \quad
\xi = \frac{\Delta \omega+\delta}{\kappa} - \frac{\Gamma}{2\Delta\omega}
\end{equation}
where $\Delta \omega$ is the frequency detuning between the photon and the dipole transition, $\delta$ is the detuning between the cavity mode and the dipole transition, $\kappa$ describes the coupling to the input and output ports and $\Gamma$ is the relaxation time of the dipole ($\Gamma = 2g^2/\kappa$). In the following we consider the case of a dipole tuned into resonance with the cavity mode ($\delta = 0$), probed with resonant light ($\Delta \omega = 0$).
If the radiation is not coupled to the dipole transition ($g=0$, $\xi \rightarrow 0$) the
cavity is transmissive, while a coupled system ($g \neq 0$, $\xi \rightarrow \infty$) can exhibit reflection of the field incident on the cavity.

\begin{figure}[ht]
\centering
\includegraphics[width=5 cm] {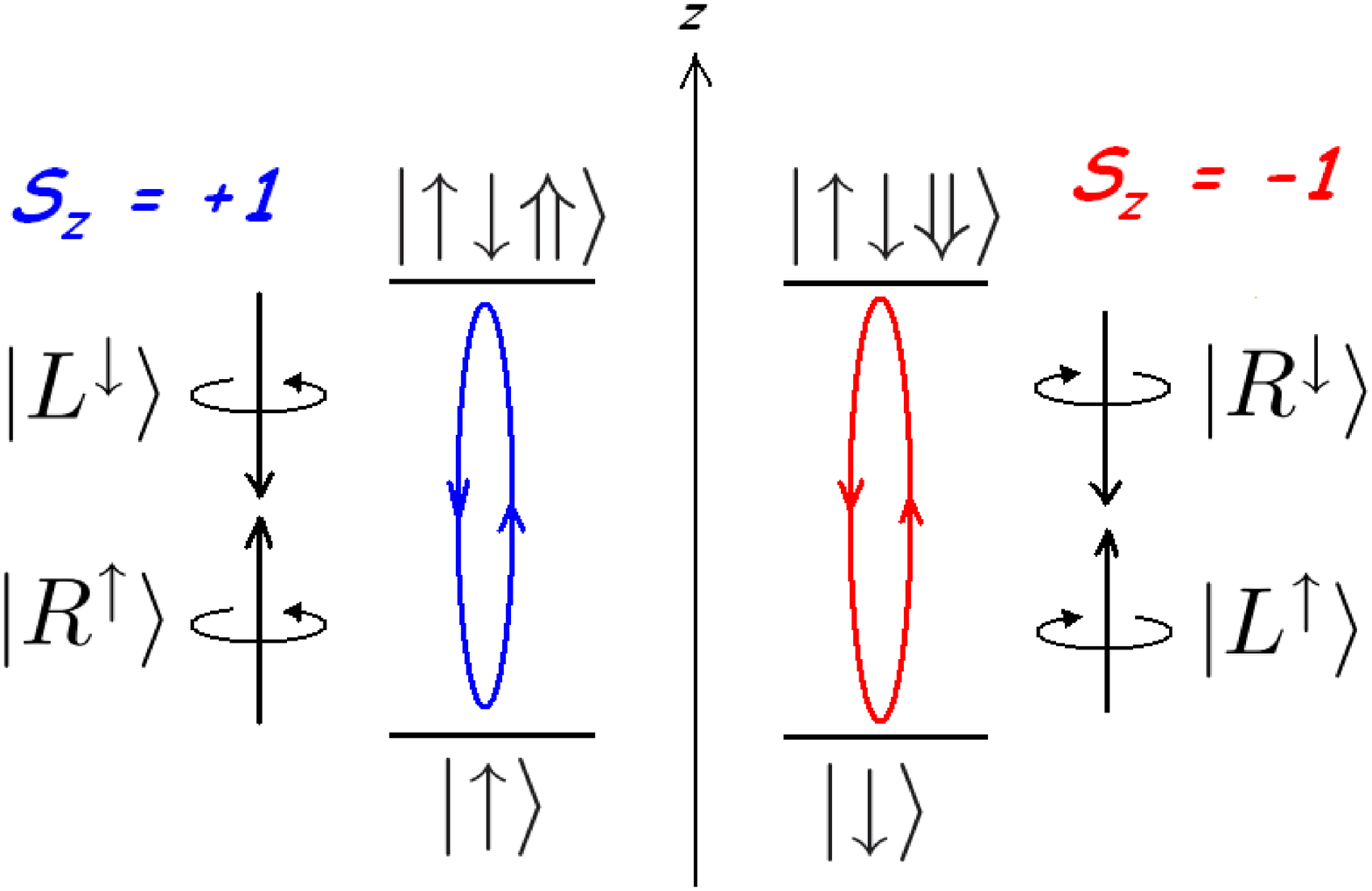}
\caption{Relevant energy levels and optical selection rules for GaAs/InAs quantum dots.}
\label{fig:notation}
\end{figure}

We now consider the dipole transitions associated with a singly charged GaAs/InAs quantum dot. The four relevant electronic levels are shown in Fig.~\ref{fig:notation}. There are two optically allowed transitions between the electron state and the trion state (bound state of two electrons and a hole). The single electron states have $J_z= \pm 1/2$ spin ($\ket{\uparrow}$, $\ket{\downarrow}$) and the holes have $J_z = \pm 3/2$ ($\ket{\Uparrow}$, $\ket{\Downarrow}$) spin. The quantization axis for angular momentum is the $z$ axis because the quantum dot confinement potential is much tighter in the $z$ (growth) direction than in the transversal direction due to the quantum dot geometry. In a trion state the two electrons form a singlet state and therefore have total spin zero, which prevents electron spin interactions with the hole spin. This makes the two dipole transitions, one involving a $s_z= +1$ photon and the other a $s_z= -1$ photon, degenerate in energy, which is a crucial requirement for achieving entanglement between photon spin and electron spin.

The spin $s_z$ of the photons in the fundamental micropillar modes is also naturally defined with respect to the z-axis.  Photon polarization is commonly defined with respect to the direction of propagation, and this causes the handedness of circularly-polarized light to change upon reflection, whereas the absolute rotation direction of its electro-magnetic fields does not change. We will therefore label the optical states by their circular polarization (labels $|L \rangle$ and $|R \rangle$) and by a superscript arrow to indicate their propagation direction along the z-axis. According to this definition, the photon spin $s_z$ remains unchanged upon reflection and the dipole-field interaction is determined only by the relative orientation of the photon spin with respect to the electron spin (see Fig.~\ref{fig:notation}).
This level scheme is idealized and does not include the non-radiative coupling between the levels, in particular due to spin
interactions with the surrounding nuclei, which lead to spin dephasing \cite{fischerSSC09}.

Consider a photon in the state $\ket{R^{\uparrow}}$ or  $\ket{L^{\downarrow}}$ ($s_z = +1$). If  the electron spin is in the state $\ket{\uparrow}$, there is dipole interaction and the photon is reflected by the cavity. Upon reflection, both the photon polarization and propagation direction are flipped and the input states are transformed respectively into the states $\ket{L^{\downarrow}}$ and $\ket{R^{\uparrow}}$. In case the electron spin is in the $\ket{\downarrow}$ state, the photon states are transmitted through the cavity and acquire a $\pi$ mod $2\pi$ phase shift relative to a reflected photon state. In the case of a $\ket{\uparrow}$ electron spin state, the interaction between the photon and the cavity is described by:
\begin {equation}
\begin{split}
\ket{R^{\uparrow}, \uparrow} \mapsto \ket{L^{\downarrow}, \uparrow} &\qquad \ket{L^{\downarrow}, \uparrow} \mapsto \ket{R^{\uparrow}, \uparrow}  \\
 \ket{R^{\downarrow}, \uparrow} \mapsto - \ket{R^{\downarrow}, \uparrow} &\qquad \ket{L^{\uparrow}, \uparrow} \mapsto -\ket{L^{\uparrow}, \uparrow},
\end{split}
\end{equation}
In the same way, the states $\ket{R^{\downarrow}}$ and $\ket{L^{\uparrow}}$ ($s_z = -1$) are reflected if the electron spin state is $\ket{\downarrow}$ and are transmitted through the cavity when the spin is $\ket{\uparrow}$.

A first application of the cavity-induced photon-spin electron-spin interface is the conditional preparation of either the $\ket{\uparrow}$ or $\ket{\downarrow}$ electron spin state. Suppose that a $\ket{R^{\uparrow}}$ photon is incident on the cavity and the electron spin is in the state $\ket{\psi_{el}} = \alpha \ket{\uparrow}+\beta \ket{\downarrow}$. Through the interaction we obtain the entangled state $\ket{\psi} = \alpha \ket{L^{\downarrow}, \uparrow} - \beta \ket{R^{\uparrow}, \downarrow}$. The detection of a photon reflected (transmitted) by the cavity projects the electron spin onto the $\ket{\uparrow}$ ($\ket{\downarrow}$) state. Electron spin projection along the $x$ or $y$ axis is not possible using photons propagating along the $z$ axis.

\begin{figure}[ht]
\centering
\includegraphics[width=8 cm] {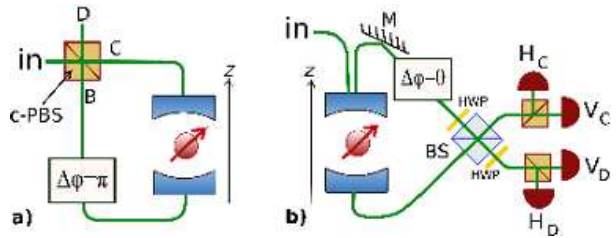}
\caption{(a) Scheme for \textsc{cnot} gate. (b) Scheme for electron spin-assisted photonic Bell-state analysis. }
\label{fig:cnot}
\end{figure}

Figure~\ref{fig:cnot}(a) shows how the interface can be used to construct a CNOT gate with the control bit the spin of the electron and the target bit the spin of the photon. Consider an incident photon in the polarization state $\ket{\psi_{ph}} = \alpha \ket{R} + \beta \ket{L}$ and an electron spin in the state $\ket{ \psi_{el}} = \gamma \ket{\uparrow} + \delta \ket{\downarrow}$. The polarizing beamsplitter in the circular basis (c-PBS) separates the input photon state into $\alpha \ket{R^{\downarrow}}$, propagating in mode $C$, and $\beta \ket{L^{\uparrow}}$, propagating in mode $B$. Eventually all photon components, either transmitted or reflected by the cavity, end up in output port D due to the polarization-flip on reflection and the properties of the c-PBS. The circuit in Fig.~\ref{fig:cnot} transforms the input state $\ket{\psi}_{\rm in} = \ket{ \psi_{ph}} \otimes \ket{ \psi_{el}}$ into:
\begin {equation} \label{eq:transformation}
\ket{\psi}_{\rm out} = \gamma \ket{\uparrow} \left[ \alpha \ket{R} + \beta \ket{L} \right] + \delta \ket{\downarrow} \left[ \alpha \ket{L} + \beta \ket{R} \right], \\
\end{equation}
provided that the phase differences in the four possible optical trajectories are equal mod$2\pi$. To this end a $\pi$ phase shift has to be included in one arm so that the two photon trajectories passing through the cavity  (in opposite directions) pick up a $\pi$ phase relative to the two possible reflective trajectories. Together with the intrinsic $\pi$ phase shift upon cavity transmission, all trajectories are in phase in the output port of the c-PBS (note that a PBS can always be constructed such that no relative phase shifts between reflected and transmitted components occur). Each arm needs to comprise an even number of mirrors, so that no net flip of polarization handedness results.
Eq.~(\ref{eq:transformation}) shows that the circuit operates as a CNOT gate, where the target photon state remains unaltered when the control electron spin is $\ket{\uparrow}$, and flips if the electron spin is $\ket{\downarrow}$.

The CNOT gate, a universal quantum gate providing entanglement between target and control qubit, has numerous applications in the field of quantum information science \cite{barencoPRL95}. For example, it can be used to mediate entangling and disentangling operations on two or more photons. Suppose the electron spin state is prepared in the $1/\sqrt{2} (\ket{\uparrow}+\ket{\downarrow}) $ state, and two uncorrelated photons, in the factorizable state $\ket{ \psi_0} =  \left( \alpha_1 \ket{R_1} + \beta_1 \ket{ L_1} \right) \left( \alpha_2 \ket{R_2} + \beta_2 \ket{L_2} \right)$, are sent through the input port one after another. After interaction with the CNOT gate, both photons will emerge in succession through the output port D, in the state:

\begin {equation}
\begin {split}
\ket{\psi} &= \ket{+} \lbrace (\alpha_1 \alpha_2 + \beta_1 \beta_2) \ket{\varphi_+} + (\alpha_1 \beta_2 + \beta_1 \alpha_2) \left | \psi_+ \right \rangle \rbrace + \\
&  \qquad \left | - \right \rangle \lbrace (\alpha_1 \alpha_2 - \beta_1 \beta_2) \left | \varphi_- \right \rangle + (\alpha_1 \beta_2 - \beta_1 \alpha_2) \left | \psi_- \right \rangle \rbrace
\end{split}
\end {equation}
where $\ket{\psi^{(\pm)}}$ and $\ket{\varphi^{(\pm)}}$ are the Bell states:
\begin{equation}
\begin {split}
\label{eq:bell}
\ket{ \varphi^{(\pm)}} = \frac{1}{\sqrt{2}} \left[ \ket{ R_1} \ket{R_2} \pm \ket{L_1} \ket{L_2}\right]
\\
\ket{\psi^{(\pm)}} = \frac{1}{\sqrt{2}} \left[ \ket{R_1} \ket{L_2} \pm \ket{L_1} \ket{R_2}\right]
\end{split}
\end{equation}
and $\ket{\pm} = 1/\sqrt{2} (\ket{\uparrow}\pm \ket{\downarrow}) $. This state is a three-particle entangled state and is written in the electron spin detection basis that will, for given $\alpha$'s and $\beta$'s, result in a specific two photon entangled state after the electron spin projection measurement. More photons can be entangled in order to create multi-photon entanglement. For example, feeding the gate with a stream of right-hand circularly polarized photons, and projecting the spin state on the $\left | \pm \right \rangle$ basis, after all the photons have interacted with the spin, N-photon GHZ states ($\ket{GHZ} = (1/\sqrt{2} \left[ \ket{L}^{\bigotimes N} +   \ket{R}^{\bigotimes N}  \right]) $)can be created. Such states have important applications, like quantum secret sharing and multiparty quantum networking.

We next present the scheme sketched in Fig.~\ref{fig:cnot}(b) for performing a deterministic and complete Bell-state analysis on an input of two subsequent photons. Consider first the 2-photon Bell states in Eq.~(\ref{eq:bell}). $\ket{\varphi}$-states can be distinguished from $\ket{\psi}$-states measuring two-photon correlations in the $\lbrace \ket{R} , \ket{L} \rbrace$ basis. Determining the $\pm$ sign in Eq.~(\ref{eq:bell}) would require correlation measurements in a linearized polarization $\lbrace \ket{H}, \ket{V} \rbrace$ basis, which is incompatible with the previous measurement. Our idea is to entangle the two photons to be analyzed with an electron spin such that each joint measurement result for the three-particle state can be uniquely associated to a single photonic Bell state.

Suppose the electron spin is prepared in $\ket{+}$. The two photons come in succession to the cavity and the reflected and transmitted paths are combined with equal path length on a 50/50 beam-splitter (BS). The reflected path can be separated from the input path by means of a polarization-maintaining fiber circulator. We assume the BS not to change the polarization on the reflected port: this can be implemented by the two half-waveplates (HWP) in Fig.~\ref{fig:cnot}(b). If the input two-photon state is $\ket{\psi^{(\pm)}}$, then the state at the output ports of the BS is (taking into account that reflection from the mirror M interchanges  $\ket{R}$ and $\ket{L}$):
\begin {equation}
\frac{1}{2} \lbrace i\left[\ket{\psi^{(\pm)}_{CC}}  + \ket{\psi^{(\pm)}_{DD}} \right] \ket{+} + \left[ \ket{\psi^{(\mp)}_{CD}} - \ket{\psi^{(\mp)}_{DC}} \right] \ket{-} \rbrace
\end {equation}
where $\ket{\psi^{(\pm)}_{ij}} = \frac{1}{\sqrt{2}} \left[ \ket{R_{1i}} \ket{L_{2j}} \pm \ket{L_{1i}} \ket{R_{2j}} \right]$. For an input $\ket{\varphi^{(\pm)}}$ state we obtain:
\begin {equation}
 \frac{1}{2} \lbrace \left[ \ket{\varphi^{(\mp)}_{CC}} - \ket{\varphi^{(\mp)}_{DD}}\right]  \ket{-} + i\left[ \ket {\varphi^{(\pm)}_{CD}} + \ket{\varphi^{(\pm)}_{DC}} \right] \ket{+} \rbrace
\end {equation}
In case both photons go out the same port (either $CC$ or $DD$), measuring the electron spin state we can identify whether the two-photon input state was $\ket{\psi}$-type (corresponding to spin $\ket{+}$) or $\ket{ \varphi}$-type (corresponding to spin $\ket{-}$). Measuring the two photons in the $\lbrace \ket{H}, \ket{V} \rbrace$ polarization basis, it is then possible to distinguish between $\ket{ \varphi^{(+)}}$ and $\ket{ \varphi^{(-)}}$ and between $\ket{ \psi^{(+)}}$ and $\ket{ \psi^{(-)}}$. Similar considerations are valid for the case where the photons exit the system through different ports. Therefore, each measurement result (consisting of photon $\lbrace \left | H \right \rangle , \left | V \right \rangle \rbrace$-polarization and output port for the two photons and spin on the $\lbrace \left | + \right \rangle , \left | - \right \rangle \rbrace$ basis for the electron) is univocally associated to a single photonic Bell state. The summary of the possible measurement results for each input Bell state is given in Table 1.\\
Phase stability is required in the two arms from the cavity to the BS, but no interferometric stability is needed between the two photons since their interaction is only through the electron spin.

\begin{table}
\caption {Output results for each photonic Bell state. Each result (consisting of polarization in the $\lbrace \ket{H}, \ket{V}\rbrace$ basis and output port for the photon and spin in the $\lbrace \ket{+}, \ket{-}\rbrace$ basis for the electron in the quantum dot) is univocally associated with one photonic Bell state.\\}

\begin{ruledtabular}
\begin{tabular}{|c|ccccc|}

  State & \multicolumn {5}{c}{Results} \\
  \hline
  $\ket{\psi^{(+)}}$       & $\ket{+}$: & $\ket{H_1^C, H_2^C}$ & $\ket{V_1^C, V_2^C}$ & $\ket{H_1^D, H_2^D}$ & $\ket{V_1^D, V_2^D}$ \\
                           & $\ket{-}$: & $\ket{H_1^C, V_2^D}$ & $\ket{V_1^C, H_2^D}$ & $\ket{H_1^D, V_2^C}$ & $\ket{V_1^D, H_2^C}$  \\
  \hline
  $\ket{\psi^{(-)}}$       & $\ket{+}$: & $\ket{H_1^C, V_2^C}$ & $\ket{V_1^C, H_2^C}$ & $\ket{H_1^D, V_2^D}$ & $\ket{V_1^D, H_2^D}$ \\
                           & $\ket{-}$: & $\ket{H_1^C, H_2^D}$ & $\ket{V_1^C, V_2^D}$ & $\ket{H_1^D, H_2^C}$ & $\ket{V_1^D, V_2^C}$   \\
  \hline
  $\ket{\varphi^{(+)}}$    & $\ket{-}$: & $\ket{H_1^C, V_2^C}$ & $\ket{V_1^C, H_2^C}$ & $\ket{H_1^D, V_2^D}$ & $\ket{V_1^D, H_2^D}$ \\
                           & $\ket{+}$: & $\ket{H_1^C, H_2^D}$ & $\ket{V_1^C, V_2^D}$ & $\ket{H_1^D, H_2^C}$ & $\ket{V_1^D, V_2^C}$ \\

  \hline
  $\ket{\varphi^{(-)}}$    & $\ket{-}$: & $\ket{H_1^C, H_2^C}$ & $\ket{V_1^C, V_2^C}$ & $\ket{H_1^D, H_2^D}$ & $\ket{V_1^D, V_2^D}$ \\
                           & $\ket{+}$: & $\ket{H_1^C, V_2^D}$ & $\ket{V_1^C, H_2^D}$ & $\ket{H_1^D, V_2^C}$ & $\ket{V_1^D, H_2^C}$ \\

\end{tabular}
\end{ruledtabular}
\end{table}

A performance parameter for a realistic system is the difference $\Delta$ between the transmission for the uncoupled and coupled cavity. From \cite{auffevesPRA07}, in the simple case of no exciton dephasing and assuming the dipole leak to be equal to its emission rate in vacuum:
\begin {equation}
\Delta = T_{max}-T_{min} = \left( \frac{Q}{Q_0} \right)^2 \left[1-\left( \frac{1}{1+F_P} \right)^2  \right]
\end{equation}
where $Q_0$ is the quality factor of the cavity due to the output coupling, $Q$ is the cavity quality factor including the leaks ($Q \leq Q_0$) and $F_P$ is the Purcell factor of the two-level system. For a micropillar cavity with oxide apertures \cite{stoltzAPL05} the optical losses due to radiation ($\alpha_{rad} =1.7 \cdot 10^{-3} \mbox{cm}^{-1}$) and aperture scattering ($\alpha_{scat} =1.7 \mbox{cm}^{-1}$) are much smaller than the photon escape losses through the top mirror ($\alpha_{m} =13.9 \mbox{cm}^{-1}$): for these values $T_{max} = (Q/Q_0)^2 \sim 0.8$. Purcell factors around $F_P = 6$ can be reached with these cavities \cite{rakherPRL09}, for which $\Delta \sim 0.78$. In general the value of $\Delta$ can be increased reducing the cavity losses and increasing the Purcell factor and the dipole lifetime.
Oxide-apertured micropillar cavities also have a very high coupling efficiency between light and the quantum dot \cite{rakherPRL09}, can incorporate intra-cavity electron charging, and can be made polarization-degenerate \cite{bonatoAPL09}. Optical fibers may be glued on both sides, etching the back wafer substrate to reduce losses. Other kind of microcavities, like photonic crystals and microdisks, can be considered as well, but light coupling is in general inefficient and polarization-degeneracy is extremely difficult to achieve, due to the intrinsic anisotropy of such structures.

A crucial aspect is the preparation of electron spin superpositions ($\ket{\pm}$). Significant progress has been made in the manipulation of single electron spins \cite{berezovskyScience08, xuNAtPhys08, pressNature08, clarkPRL09}. Spin manipulation typically requires Zeeman splitting of the spin ground states, which may be achieved with a magnetic field or through optical Stark effect. Ground state degeneracy, with Zeeman splitting less than photon bandwidth, has to be restored in the implementation of quantum information protocols. Ultrafast spin manipulation through ac-Stark effect, potentially in addition to a weak magnetic field (as shown in \cite{berezovskyScience08}), seems more promising for our purposes than any preparation involving strong magnetic fields, whose modulation is extremely challenging on timescales shorter than the spin coherence time.\\
Quantum optical applications, like photon entangling gate and BSA, require the phase of spin superposition to be constant at the times of interaction with different photons. The dephasing time is typically around $5-10$ ns \cite{pettaScience05, berezovskyScience08} but can be increased by several orders of magnitude by spin echo techniques and manipulations of the nuclear spins \cite{oultonPRL07, greilichScience07, reillyScience08, vinkNatPhys09, lattaNatPhys09, clarkPRL09}.

Finally we point out that the combination of conditional spin-preparation and probing based on spin-state selective reflection could be used to investigate the dynamics of the quantum dot electron spin state \cite{Bouwmeester2009}.

In conclusion, we introduced a quantum interface between a single photon and the spin state of an electron trapped in quantum dot, based on cavity-QED in the weak-coupling regime. We proposed as possible applications: a spin-photon CNOT gate, a multi-photon entangled state generator, and a photonic Bell-state analyzer.

We thank M. Rakher and D. Loss for stimulating discussions. This work was supported by the NSF grant 0901886, and the Marie-Curie No. EXT-CT-2006-042580. C.B., F. H. and S. S. R. O. contributed equally to this work.

\end{document}